\documentclass[journal,comsoc]{IEEEtran}
\usepackage[T1]{fontenc}
\usepackage{multirow}
\usepackage[bookmarks=false]{hyperref}
\usepackage{longtable}
\usepackage{supertabular} 
\usepackage{graphicx}
\usepackage{lineno,hyperref}
\usepackage{amsmath}
\usepackage{amsfonts}
\usepackage{amssymb}
 \usepackage{flushend}
\usepackage{array,multirow,makecell}
\modulolinenumbers[5]
\usepackage{lscape}
\usepackage[cmintegrals]{newtxmath}
\usepackage[table]{xcolor}

\IEEEoverridecommandlockouts                         

\title{Authentication and Authorization for Mobile IoT Devices using Bio-features: Recent Advances and Future Trends}

\author{Mohamed Amine Ferrag$^{1}$, Leandros Maglaras$^{2}$, Abdelouahid Derhab$^{3}$
\thanks{This paper was presented in part at the 3rd International Conference on Pattern Recognition and Intelligent Systems (PAIS 2018), 24-25 October 2018, Tebessa, Algeria}
\thanks{$^{1}$ Department of Computer Science, Guelma University, B.P. 401, 24000, Algeria e-mail: mohamed.amine.ferrag@gmail.com, ferrag.mohamedamine@univ-guelma.dz}
\thanks{$^{2}$ School of Computer Science and Informatics, De Montfort University, Leicester, UK, , and also with General Secretariat of Digital Policy, Athens, Greece, e-mail:  leandros.maglaras@dmu.ac.uk}
\thanks{$^{3}$ Center of Excellence in Information Assurance (CoEIA), King Saud University, Saudi Arabia, e-mail: abderhab@ksu.edu.sa}
}
\begin{document}
\maketitle
\thispagestyle{empty}
\pagestyle{empty}
\begin{abstract}
Bio-features are fast becoming a key tool to authenticate the IoT devices; in this sense, the purpose of this investigation is to summaries the factors that hinder biometrics models' development and deployment on a large scale, including human physiological (e.g., face, eyes, fingerprints-palm, or electrocardiogram) and behavioral features (e.g., signature, voice, gait, or keystroke). The different machine learning and data mining methods used by authentication and authorization schemes for mobile IoT devices are provided. Threat models and countermeasures used by biometrics-based authentication schemes for mobile IoT devices are also presented. More specifically, We analyze the state of the art of the existing biometric-based authentication schemes for IoT devices. Based on the current taxonomy, We conclude our paper with different types of challenges for future research efforts in biometrics-based authentication schemes for IoT devices.
\end{abstract}

\section{Introduction}

Biometric identification enables end-users to use physical attributes instead of passwords or PINs as a secure method of accessing a system or a database. Biometric technology is based on the concept of replacing 'one thing you have with you' with 'who you are', which has been seen as a safer technology to preserve personal information. The possibilities of applying biometric identification are really enormous. 

Biometric identification is applied nowadays in sectors where security is a top priority, like airports and could be used as a means to control border crossing at sea, land and air frontier \cite{del2016automated}. Especially for the air traffic area, where the number of flights will be increased by 40 \% before 2013, the authentication of mobile IoT devices will be achieved when the bio-features models becomes sufficiently mature, efficient and resistant to IoT attacks.

Another area where biometric identification methods are starting to be adopted is electronic IDs. Biometric identification cards such as the Estonian and Belgian national ID cards were used in order to identify and authenticate eligible voters during elections. Moving one step further, Estonia has introduced the Mobile-ID system that allows citizens to conduct Internet voting \cite{vinkel2016and}, that combines biometric identification and mobile devices. This system which was quite innovative when it was initialy introduced posses several threats to the electoral procedure and was criticized for being insecure \cite{springall2014security}. 
\begin{figure} \label{fig:Fig1}
\centering
\includegraphics[width=0.8\linewidth]{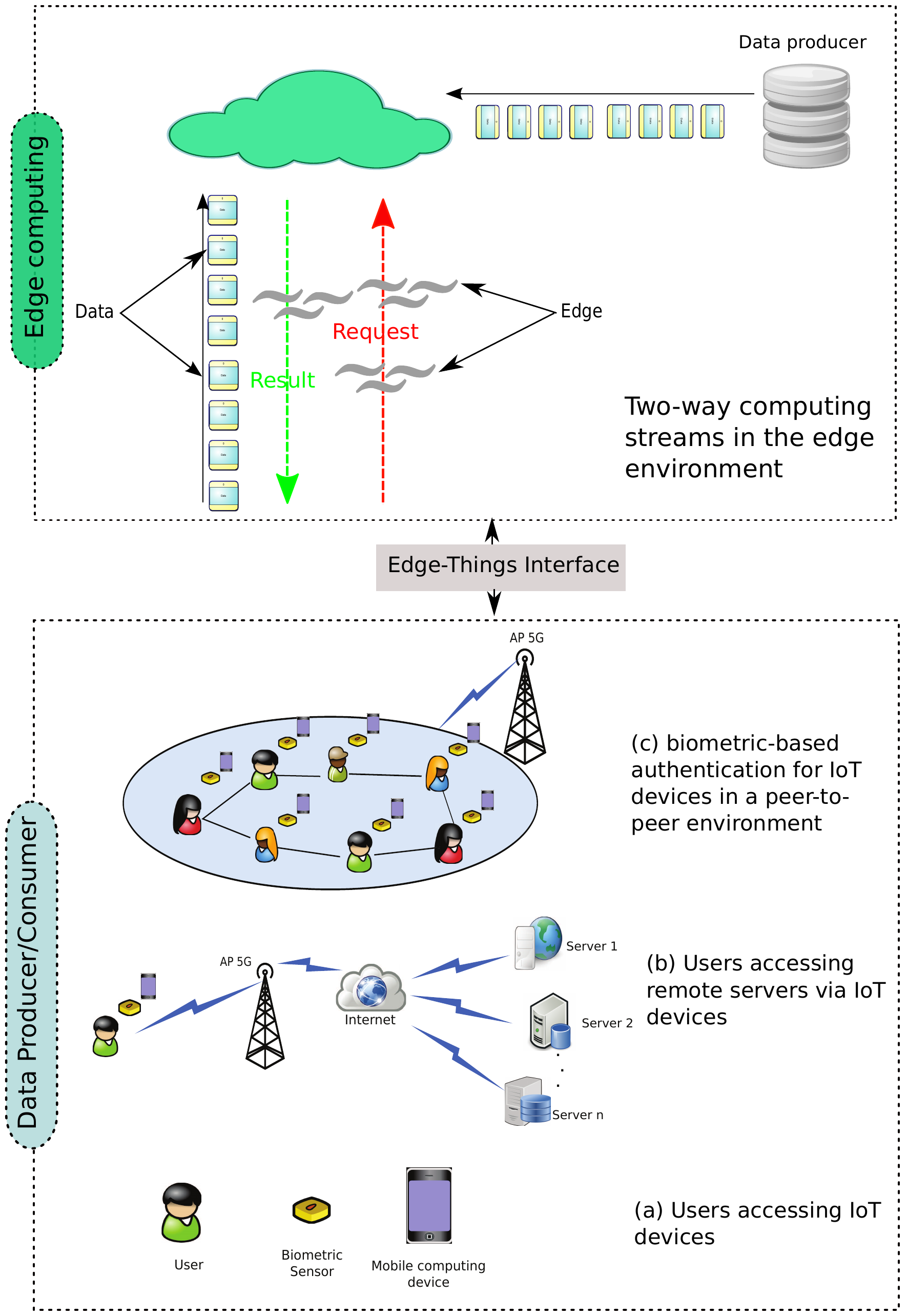}
\caption{Types of communication for IoT devices in edge environments during the authentication and authorization, (a) users accessing IoT devices, (b) users accessing remote servers via IoT devices, (c) biometric-based authentication for IoT devices in a peer-to peer environment}
\label{fig:Figa}
\end{figure}
\begin{table*}[ht]
\centering{
    \caption{Related surveys on Biometric Authentication} \label{Table:survey}
\begin{tabular}{l|c|c|c|c|c}
  \hline
  \rowcolor{black!15}\textbf{Reference} &  \textbf{Deployment Scope}  & \textbf{Focus Biomtetric Area}  & \textbf{Threat models} & \textbf{Countermeasures} & \textbf{ML and DM} \\
  \hline
  Gafurov (2007) \cite{gafurov2007survey}  & Not mobile & Gait recognition & No  & No & No \\  \hline
  \rowcolor{black!5}Revett et al. (2008) \cite{revett2008survey}   & Not mobile  & Mouse dynamics & No & No & No\\  \hline
  Yampolskiy and Govindaraju (2008) \cite{yampolskiy2008behavioural}  & Not mobile  & Behavioral-based & No  & No  & No\\  \hline
  \rowcolor{black!5}Shanmugapriya and Padmavathi (2009) \cite{shanmugapriya2009survey} & Not mobile & Keystroke dynamics & No  & No  & Yes\\  \hline
  Karnan et al. (2011) \cite{karnan2011biometric} & Not mobile &  Keystroke dynamics & No & No  & Yes\\  \hline
  \rowcolor{black!5}Banerjee and Woodard (2012) \cite{banerjee2012biometric} & Not mobile &  Keystroke dynamics & No & No & Yes\\  \hline
  Teh et al. (2013) \cite{teh2013survey} & Not mobile  &  Keystroke dynamics & No & No &  Yes\\  \hline
  \rowcolor{black!5}Bhatt et al. (2013) \cite{bhatt2013keystroke} &  Not mobile  &  Keystroke dynamics & No & No & Yes\\  \hline
  Meng et al. (2015) \cite{10} & Mobile device & All  & Yes & Yes & Partial\\  \hline
  \rowcolor{black!5}Teh et al. (2016) \cite{6} & Mobile device & Touch dynamics  & No & No & Yes\\  \hline
  Mahfouz et al. (2017) \cite{mahfouz2017survey}  & Smartphone &  behavioral-based & No & No & Yes\\  \hline
  \rowcolor{black!5}Mahadi et al. (2018) \cite{mahadi2018survey} & Not mobile & behavioral-based &  No &  No & Yes\\  \hline
  Sundararajan and Woodard (2018) \cite{sundararajan2018deep} & Not mobile & All & No & No  & Yes\\  \hline
  \rowcolor{black!5}Rattani and Derakhshani (2018) \cite{rattani2018survey} & Mobile device & Face recognition & Yes & Yes &  Yes \\  \hline
  Our survey &  Mobile IoT device & All & Yes & Yes & Yes\\
    \hline
\end{tabular}\\}
ML and DM: Machine learning (ML) and data mining (DM) algorithms
\end{table*}
According to a survey by Javelin Strategy \& Research, in 2014, \$16 billion was stolen by 12.7 million people who were victims of identity theft, only in the US \cite{sen2015estimating}. This amount is calculated without taking into account the economic problems and psychological oppression that victims of this fraud suffer. From the banking sector and businesses, to access to homes, cars, personal computers and mobile devices, biometric technology offers the highest level of security in terms of privacy and privacy protection and secure access.

Mobile devices are nowadays an essential part of our everyday life, as they are used for a variety of mobile applications. Performing biometric authentication through mobile devices can provide a stronger mechanism for identity verification as the two authentication factors: "something you have" and "something you are" are combined. Several solutions that include multi-biometric and behavioral authentication platforms for telecom carriers, banks and other industries were recently introduced \cite{unitedbiometrics}.

In the literature, many authentication schemes based on bio-features models for mobile IoT devices have been proposed. As shown in Figure \ref{fig:Fig1}, the schemes can perform two different authentication operations: they either (a) authenticate the users to access the mobile devices,  or (b) authenticate the users to access remotes servers through mobile devices. The main challenges that are facing biometric-based authentication schemes are: (1) how to design an authentication mechanism that is free from vulnerabilities, which can be exploited by adversaries to make illegal accesses, and (2) how to ensure that the user's biometric reference templates are not compromised by a hacker at the device-level or the remote server-level.

Our contributions in this work are:

\begin{itemize}

\item We classify the related surveys according to several criteria, including, deployment Scope, focus biometric area, threat models, countermeasures, and ML/DM algorithms.

\item We present the machine learning and data mining methods used by authentication and authorization schemes for mobile IoT devices, including, unsupervised, semi-supervised, and supervised approaches.

\item We present all the Bio-features used by authentication and authorization schemes for mobile IoT devices.

\item We provide a comprehensive analysis and qualitative comparison of the existing authentication and authorization schemes for mobile IoT devices.

\item We emphasize the challenges and open issues of authentication and authorization schemes for mobile IoT devices.
\end{itemize}

The rest of this paper is organized as follows. Section \ref{sec:survey} gives the related surveys on biometric authentication. In Section \ref{sec:machine}, we present the different machine learning and data mining algorithms used by authentication and authorization schemes for mobile IoT devices. In Section \ref{sec:conter}, we provide the new trends of biometric technologies including human physiological (e.g., face, eyes, fingerprints-palm, and electrocardiogram) and behavioral (e.g., signature, voice, gait, or keystroke). In Section \ref{sec:bio}, we clearly highlight the pros and cons of the existing authentication and authorization schemes for mobile IoT devices. Then, we discuss the challenges and suggest future research directions in both Section \ref{sec:future} and \ref{sec:future}. Lastly, Section \ref{sec:con} presents conclusions.

\section{Related surveys on biometric authentication}\label{sec:survey}
In the literature, there are different related surveys that deal with user authentication. Although some of them covered different authentication methods \cite{5, aslam2017survey, kunda2018survey}, but we only consider those that were fully dedicated for biometric authentication. We classify the surveys according to the following criteria:

\begin{itemize}
       \item \textit{Deployment Scope:} It indicate whether the authentication scheme is deployed on mobile devices or not.
        \item \textit{Focus biometric area:} It indicates whether the survey focused on all/specific biometric features.
        \item \textit{Threat models:} It indicates whether the survey considered the threats against the authentication schemes.
        \item \textit{Countermeasures:} It indicates whether the survey focused considered the countermeasures to defend the authentication schemes.
        \item \textit{Machine learning (ML) and data mining (DM) algorithms:} It indicates whether the survey mentions for each solution the used machine learning or data mining method. 
\end{itemize}

Some surveys described the authentication schemes that only consider specific bio-features. For instance, the surveys \cite{shanmugapriya2009survey, karnan2011biometric, banerjee2012biometric, bhatt2013keystroke, teh2013survey} only focused on the keystroke dynamics. On the other hand, Gafurov \cite{gafurov2007survey} presented biometric gait recognition systems. Revett et al. \cite{revett2008survey} surveyed biometric authentication systems that rely on mouse movements. Yampolskiy and Govindaraju \cite{yampolskiy2008behavioural} presented a a comprehensive study on behavioral biometrics. Mahadi et al. \cite{mahadi2018survey} surveyed behavioral-based biometric user authentication, and determined the set of best classifiers for behavioral-based biometric authentication. Sundararajan and Woodard \cite{sundararajan2018deep} surveyed different 100 approaches that leveraged deep learning and various biometric modalities to identify users. Teh et al.  \cite{6} presented different authentication solutions that rely on touch dynamics in mobile devices. Rattani and Derakhshani \cite{rattani2018survey} provided the state-of-the-art related to face biometric authentication schemes that are designed for mobile devices. They also discussed the spoof attacks that target mobile face biometrics as well as the anti-spoofing methods. Mahfouz et al. \cite{mahfouz2017survey} surveyed the behavioral biometric authentication schemes that are applied on smartphones. Meng et al. \cite{10} surveys the authentication frameworks using biometric user on mobile phones. They identified eight potential attack against these authentication systems  along with promising countermeasures. Our survey and \cite{10} both focus on authentication schemes that are designed for mobile device, and consider all the biometric features, and deal with threat models and countermeasures. However, \cite{10} do not give information related to the used machine learning or data mining method of all the surveyed solutions. In addition, \cite{10} only covers papers up to 2014, whereas the coverage of our survey is up to 2018. To the best of our knowledge, this work is the first that thoroughly covers threats, models, countermeasures, and the machine learning algorithms of the biometric authentication schemes.

\begin{table}[!t]
\scriptsize
	\centering
	\caption{Machine learning and data mining methods used by authentication and authorization schemes for mobile IoT devices}
	\label{Table:Tabmachine}
	\setlength{\tabcolsep}{2pt}
		\vspace*{-\baselineskip}
		\renewcommand{\arraystretch}{1.5}
{
\begin{tabular}{p{1.8in}|p{1.4in}} \hline 
\rowcolor{black!15}\textbf{Machine learning and data mining methods} & \textbf{Schemes} \\ \hline \hline  
Agglomerative complete link clustering approach & \cite{31} \\ \hline 
Support vector distribution estimation & \cite{90} \cite{39} \\ \hline 
\rowcolor{black!5}Gaussian mixture model & \cite{15} \\ \hline 
Embedded hidden Markov model & \cite{15} \\ \hline 
\rowcolor{black!5}k-nearest-neighbors (kNN) & \cite{39} \cite{lin2012new} \cite{shen2012continuous} \cite{jagadeesan2009novel} \cite{buriro2019answerauth}\\ \hline 
Support-vector machine (SVM) & \cite{39} \cite{72} \cite{63} \cite{70} \cite{F7} \cite{50} \cite{N7} \cite{lin2012new} \cite{shen2012continuous} \cite{zheng2011efficient} \cite{bailey2014user} \cite{fridman2015multi} \cite{buriro2019answerauth}\\ \hline 
\rowcolor{black!5}A computation efficient statistical classifier & \cite{47} \\ \hline
Deep learning & \cite{N1} \cite{das2018deep} \cite{F9} \cite{bayar2016deep} \cite{alhussein2018voice}\\ \hline 
\rowcolor{black!5}Local binary patterns algorithm & \cite{79} \\ \hline 
Mel frequency cepstral coefficients & \cite{N2} \\ \hline 
\rowcolor{black!5}Pupillary light reflex & \cite{N3} \\ \hline 
Euclidean distance, hamming distance & \cite{N4} \\ \hline 
\rowcolor{black!5}Deep convolutional neural network & \cite{63} \cite{liu2018finger} \cite{ranjan2018deep} \cite{rattani2018multi}\\ \hline 
Genetic algorithm & \cite{khalifa2018multimodal} \\ \hline 
\rowcolor{black!5}Artificial neural network (ANN) & \cite{N7} \\ \hline 
Gauss-newton based neural network & \cite{N6} \\ \hline 
\rowcolor{black!5}Radial integration transform & \cite{N8} \\ \hline 
Weibull distribution  & \cite{gamboa2007webbiometrics} \\ \hline 
\rowcolor{black!5} Online learning algorithms & \cite{cai2018online} \\ \hline
Counter-Propagation Artificial Neural Network (CPANN) & \cite{N7}\\ \hline 
\rowcolor{black!5} Random Forest (RF)  & \cite{feher2012user} \\ \hline
Neural Network (NN) & \cite{ahmed2014biometric} \cite{shen2012continuous} \cite{jagadeesan2009novel}\\ \hline
Circular integration transform & \cite{N8} \\ \hline
\rowcolor{black!5} Decision Tree (DT) & \cite{lin2012new} \cite{bailey2014user} \cite{sheng2005parallel} \cite{patel2015improving} \cite{kumar2013fuzzy} \cite{buriro2019answerauth}\\ \hline
Learning Algorithm for Multivariate Data Analysis (LAMDA)
 & \cite{nakkabi2010improving} \\ \hline 
\rowcolor{black!5}  Bayesian network (BN) & \cite{bailey2014user} \cite{buriro2019answerauth} \\ \hline 
Naive Bayes & \cite{fridman2015multi} \cite{traore2012combining} \cite{buriro2019answerauth}
\\
\rowcolor{black!5} Pearson product-moment correlation coefficient (PPMCC)  & \cite{jagadeesan2009novel} \\ \hline 
Keyed random projections and arithmetic hashing & \cite{N5} \\ \hline 
\rowcolor{black!5} One-dimensional multi-resolution local binary patterns & \cite{N9} \\ \hline
\end{tabular}}
\end{table}

\section{Machine learning and data mining algorithms }\label{sec:machine}

In this section, we lists the different machine learning and data mining algorithms used by biometric-based authentication schemes for IoT devices, as presented in \ref{Table:Tabmachine} .

\subsection{Support-vector machine (SVM)}
The SVM is a popular and powerful binary classifier, which aims to find a hyperplane within the feature space that separates between two classes. SVM is used by seven authentication schemes for IoT devices in edge environments using bio-features \cite{39, 72, 63, 70, F7, 50, N7}.

In \cite{39}, Frank et al. used two classifiers: K-Nearest-Neighbors (kNN), and SVM with an RBF-kernel. In this study, two classes are chosen, namely, i) user of interest and ii) the rest of users. In the training data phase, this study tune the two relevant parameters, i.e., $\gamma $ and $C$ of the RBF-SVM, are tuned under five-fold cross-validation. The first parameter $\ \gamma $ is used for controlling the gaussian radial-basis function. The second parameter $C$ is used for controlling the trade-off between maximizing the margin and minimizing the number of exceptions.

In Sitova et al. \cite{72}, an SVM classifier with scaled Manhattan (SM) and scaled Euclidian (SE) are used to perform verification experiments. For parameter tuning, the RBF kernel was selected to perform a grid search to find the parameter.

In order to detect faces of a particular size, Sarkar et al. \cite{63} introduced a face detection algorithm, wich is based on deep feature combined with a SVM classifier. Specifically, the study passes the image through a deep convolutional neural network, then they used train SVMs of different sizes in order to achieve scale invariance. Durang training step, Sarkar et al.'s scheme uses 5202 images from the UMD-AA database, which is a database of 720p videos and touch gestures of users on a mobile device (iPhone). The experimental results showed that the proposed idea can detect the partial or the extremely posed faces in IoT environment.

The approach described by Mahbub et al. \cite{70} is a framework for authentication and authorization of users' faces on Mobile IoT devices. Their approach trains a linear SVM with statistical features. The study used the Active Authentication Dataset, which contains the front-facing camera face video for 50 iPhone users (43 male, 7 female) with three different ambient lighting conditions, including, well-lit, dimly-lit, and natural daylight. Compared to Viola-Jones face detector, the Mahbub et al.'s framework can achieve superior performance.

In another study, the SVM classifier was attempted as the learning algorithm by Gunasinghe and Bertino \cite{F7}, face as the bio-feature
, and eigen faces as the feature extraction algorithm. The trained SVM classifier helps to the artifacts stored in the Mobile IoT devices. Compared to Mahbub et al.'s \cite{70} approach, the protocol \cite{F7} considers privacy preserving of the training data, which is uses three secrets $(S_i:\ i\ \in \ \{1,\ 2,\ 3\})$ in different phases of the scheme, including, $S_1$ of size $128\ bits$, $S_2$ of size $160\ bits$, and $S_3$ of size $256\ bits$.

Chen et al. \cite{50} introduced a two-factor authentication protocol using rhythm, which can be applied for mobile IoT devices. Specifically,  Chen et al.'s protocol employs SVM as a machine learning classifier, and LibSVM in the implementation phase. The experimental results on Google Nexus 7 tablets, involving 22 legitimate users and 10 attackers, show an outstanding results. The false-positive and false-negative rates achieve 0.7\% and 4.2\%, respectively. In general, there are two behavioral biometric modalities in the construction of an authentication scheme based on the bio-feature, including, 1) Using one behavioral biometric model, which does not need any additional hardware to capture data, and 2) Using a combination of the behavioral biometric models.

\subsection{Deep learning approach}
Actually, Deep learning is used to authenticate low-power devices in the IoT networks. Deep Learning approach is based on an artificial neural network (ANN), consisting of many layers of neurons, referred to as hidden layers, between two other layers: input and output. Each layer receiving and interpreting information from the previous layer. Unlike SVM, the learning runtime increases when the number of features in an ANN increases. Ferdowsi and Saad \cite{N1} proposed a deep learning method based on the long short-term memory (LSTM), which uses the fingerprints of the signal $y$ generated by an IoT mobile device. In addition,  LSTM algorithm is used to allow an IoT mobile device updating the bit stream by considering the sequence of generated data. The paper expressed that the findings were reported that dynamic LSTM watermarking is able to detect some attacks such as eavesdropping.

Das et al. \cite{das2018deep} used a deep-learning based classifier to have a faster system against high-power adversaries. Similarly to the work \cite{N1}, this study uses the long short-term memory (LSTM). The experiments used a testbed of LoRa low-power wireless, which consists of 29 Semtech SX1276 chips as LoRa transmitters and a Semtech SX1257 chip as the receiver. The experimental results showed that the classification performance is more promising with respect to state-of-the-art LoRa transmitters.

The work by Bazrafkan and Corcoran \cite{38} used a deep U-shaped network with 13 layers for the segmentation task. The study used a 3x3 kernel that maps the input to the first convolutional hidden layer in order to enhance iris authentication for Mobile IoT devices. They used two databases, including, 1) CASIA Thousand, which contains 20k images, and 2) Bath 800, which contains 24156 images. The segmentation results are reported as 98.55\% for the Bath 800 and 99.71\% for CASIA Thousand. The paper also states the benefits of the deep learning technique such as efficient segmentation on large data sets.

In their study, Bayar and Stamm \cite{bayar2016deep} use a universal forensic approach using deep learning in order to detect multiple types of image forgery. For image recognition, the convolutional neural networks (CNNs) is used as tool from deep learning. Specifically, the CNN proposed contains eight layers, including, the proposed new convolutional layer, two convolutional layers, two max-pooling layers, and three fully-connected layers. The first layer of the network is 227 $\times$ 227 grayscale image. The proposed CNN is evaluated as a binary and multi-class classifier. Although the false positive rate is not reported, the Caffe deep learning framework is used, which shows that the CNN proposed model can distinguish between unaltered and manipulated images with at least 99.31\% and 99.10\% accuracy for a binary and multi-class classifier, respectively.

\subsection{Deep convolutional neural network}

The deep convolutional neural networks (DCNNs) for face detection was attempted by Ranjan et al. \cite{ranjan2018deep}, which can be classified into two categories, including, the region-based approach and the sliding-window approach. The DCNN can identify whether a given proposal contains a face or not.

Based on deep learning and random projections, Liu et al. \cite{liu2018finger} proposed a novel finger vein recognition algorithm, named FVR-DLRP, which could be used for Mobile IoT devices. The FVR-DLRP algorithm uses four main phases, namely, 1) feature extraction, 2) random projection, 3) training, and 4) matching. The finger vein feature extraction is based on $3\times 3$ regions. The Johnson--Lindenstrauss theorem is used for the random projections. In the training phase, the Deep belief network is applied to generating the biometric template. The experimental results on finger vein laboratory database, named FV\_NET64, involving 64 people's finger vein image, and each of them contributes 15 acquisitions, show that the FVR-DLRP algorithm achieves 91.2\% for recognition rate (GAR) and 0.3\% for false acceptance rate (FAR). In the study by Sarkar et al. \cite{63}, a deep convolutional neural network is proposed for mobile IoT devices. According to the study, the OpenCL and RenderScript based libraries for implementing deep convolutional neural networks are more suitable for mobile IoT devices compared to the CUDA based schemes.

\subsection{Decision Tree (DT)}
DTs are a type of learn-by-example pattern recognition method, which were used by five studies \cite{sheng2005parallel} \cite{lin2012new} \cite{kumar2013fuzzy} \cite{bailey2014user} \cite{patel2015improving}. In \cite{sheng2005parallel}, Sheng et al. proposed a parallel decision trees based-system in order to authenticate users based on keystroke patterns, which could be applied for mobile IoT devices. According to the study, a parallel DT alone cannot solve the authentication on keystroke patterns. The training data contains 43 users, each of them typed a given common string of 37 characters. The study achieves 9.62\% for FRR and 0.88\% for FAR. Therefore, Kumar et al. \cite{kumar2013fuzzy} presented a fuzzy binary decision tree algorithm, named FBDT, for biometric-based personal authentication. The FBDT was able to detect with FAR=0.005\% and FRR=3.027\% on palmprint, and FAR=0.023\% and FRR=8.1081\% on iris, and FAR=0\% and FRR=2.027\% on the bimodal system. To enhance the network authentication in ZigBee devices, Patel et al. \cite{patel2015improving} presented an authentication system  that employs ensemble decision tree classifiers. Specifically, the study applied Multi-Class AdaBoost ensemble classifiers and non-parametric Random Forest on the fingerprinting arena.

\subsection{k-nearest-neighbors (kNN)}
The kNN algorithm identifies the $k$ training observations to belong to a group among a set of groups based on a distance function in a vector space to the members of the group \cite{jagadeesan2009novel}. In our study, we found that it is always combined with other classifiers in order to provides a fast classification. The study \cite{39} uses the kNN algorithm and a support-vector machine with an rbf-kernel. The study \cite{lin2012new} combines three classifiers, namely, the kNN algorithm, support vector machines, and decision trees. The study \cite{shen2012continuous} combines three models, including, 1) a nearest-neighbor based detector model, 2) a neural network detector model, and 3) a support vector machine model. The study by Jagadeesan and Hsiao \cite{jagadeesan2009novel} incorporates statistical analysis, neural networks, and kNN algorithms, which the experimental results show that the identification accuracy is 96.4\% and 82.2\% in case of the application-based model and the the application-independent model, respectively.

\subsection{Statistical models}
In order to perform authentication of the user's identity on mobile IoT devices, Tasia et al. \cite{47} used a computation efficient statistical classifier, which has low computational complexity compared to fuzzy logic classifiers and do not require comparison with other users' samples for identification. Therefore, hidden Markov model is a statistical model where Kim and Hong \cite{15} used an embedded hidden Markov model algorithm and the two-dimensional discrete cosine transform  for teeth authentication. For the voice authentication on mobile IoT devices, the study use pitch and mel-frequency cepstral coefficients as feature parameters and a Gaussian mixture model algorithm to model the voice signal. In the experiment section, Kim's study used an Hp iPAQ rw6100 mobile device equipped with a camera and sound-recording device. The study reported an ERR of 6.42\% and 6.24\% for teeth authentication and voice authentication, respectively. 

\subsection{Naive Bayes}
To map from the feature space to the decision space, Fridman et al. \cite{fridman2015multi} used the Naive Bayes classifier, which is based on the so-called Bayesian theorem. In the experiment section, the study reached a false acceptance rate of 0.004 and a false rejection rate of 0.01 after 30 seconds of user interaction with the device. Therefore, Traore et al. \cite{traore2012combining} considered two different biometric modalities, namely, keystroke and mouse dynamics. Their study used a Bayesian network to build the user profile, and then use it to classify the monitored samples. The experimental results show that the mouse dynamics model has a reached an equal error rate (EER) of 22.41\%, which is slightly lower than the keystroke dynamics that reached an EER of 24.78\%. In addition, Bailey et al. \cite{bailey2014user} used a Bayesian network with two machine learning algorithms, including, LibSVM and J48. The results achieved a full fusion false acceptance rate of 3.76\% and a false rejection rate of 2.51\%.

To solve the problem of verifying a user, Buriro et al. \cite{buriro2019answerauth} proposed AnswerAuth, an authentication mechanism, which is based on the extracted features from the data recorded using the built-in smartphone sensors. In effect, the AnswerAuth mechanism is tested using a dataset composing of 10, 200 patterns (120 from each sensor) from 85 users and six classification techniques are used, including, Bayes network, naive Bayes, SVM, kNN, J48, and Random Forest. According to the study, Random Forest classifier performed the best with a true acceptance rate of 99.35\%.

\begin{figure}
\centering
\includegraphics[width=1\linewidth]{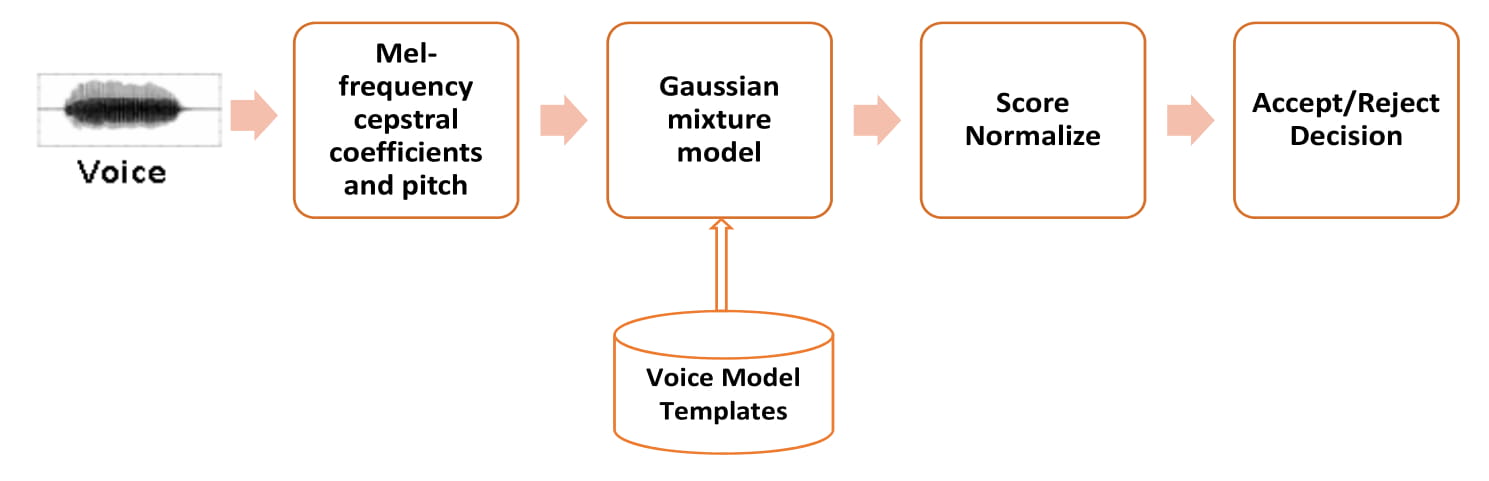}
\caption{A voice-based authentication scheme}
\label{fig:Fig2}
\end{figure}
\begin{table}[!t]
\scriptsize
	\centering
	\caption{Bio-features used by authentication schemes for IoT devices in edge environments}
	\label{Table:Tab1}
	\setlength{\tabcolsep}{2pt}
		\vspace*{-\baselineskip}
		\renewcommand{\arraystretch}{1.5}
{
\begin{tabular}{p{1.2in}|p{1.3in}} \hline \hline
\rowcolor{black!15}\textbf{Bio-feature} & \textbf{Schemes} \\ \hline \hline 
Gaze gestures & \cite{62} \cite{84} \cite{F1} \\ \hline 
\rowcolor{black!5}Electrocardiogram & \cite{61} \cite{74} \\ \hline 
Voice recognition & \cite{15} \cite{50} \cite{ali2018edge} \cite{alhussein2018voice}\\ \hline 
\rowcolor{black!5}Signature recognition & \cite{90} \\ \hline
Gait recognition & \cite{51} \\ \hline
\rowcolor{black!5}Behavior profiling & \cite{39}\cite{72}\cite{90}\cite{80} \\ \hline
Keystroke dynamics & \cite{31} \cite{47} \cite{60} \cite{ahmed2014biometric} \cite{traore2012combining} \cite{N7} \cite{sheng2005parallel}
\\ \hline
\rowcolor{black!5}Touch dynamics & \cite{6} \cite{F1} \\ \hline 
Fingerprint & \cite{14} \cite{27} \cite{32} \cite{44} \cite{F3} \cite{F4} \cite{kumar2013fuzzy}  \\ \hline 
\rowcolor{black!5}Smart card & \cite{22} \cite{28} \cite{F2} \\ \hline 
Multi-touch interfaces & \cite{35} \cite{45} \\ \hline 
\rowcolor{black!5}Graphical password & \cite{34} \\ \hline 
Face recognition & \cite{46} \cite{63} \cite{70} \cite{71} \cite{F7} \\ \hline 
\rowcolor{black!5}Iris recognition & \cite{46} \cite{53} \cite{68} \cite{F9} \\ \hline 
Rhythm & \cite{50} \\ \hline 
\rowcolor{black!5}Capacitive touchscreen & \cite{52} \\ \hline 
Ear Shape & \cite{79} \\ \hline
\rowcolor{black!5}Arm gesture & \cite{79} \\ \hline 
Plantar biometrics & \cite{F5} \\ \hline
\rowcolor{black!5} Mouse dynamics & \cite{shen2012continuous} \cite{zheng2011efficient} \cite{traore2012combining} \cite{N7} \cite{bailey2014user}
\\ \hline
Slap fingerprints & \cite{gupta2018multibiometric} \\ \hline
\rowcolor{black!5} Palm dorsal vein & 
\cite{gupta2018multibiometric} \\ \hline
\rowcolor{black!5} Hand geometry & \cite{gupta2018multibiometric} \\ \hline
Behavioral biometric &  \cite{cai2018online}\\ \hline
\end{tabular}}
\end{table}

\section{Bio-features}\label{sec:conter}

The Bio-features used by authentication and authorization schemes for mobile IoT devices can be classified in two types, including human physiological (e.g., face, eyes, fingerprints-palm, or electrocardiogram) and behavioral (e.g., signature, voice, gait, or keystroke). Tab. \ref{Table:Tab1} presents the biometrics-based authentication schemes for mobile IoT devices with Bio-features used as a countermeasure.

\begin{itemize}
\item  Gaze gestures: By combining gaze and touch, Khamis et al. \cite{62} introduced multimodal authentication for mobile IoT devices, which is more secure than single-modal authentication against against iterative attacks and side attacks.

\item  Electrocardiogram: Electrocardiogram methods can conceal the biometric features during authentication, which are classified as either electrocardiogram with the fiducial features of segmented heartbeats or electrocardiogram with non-fiducial features as discussed in \cite{61} \cite{74}. Both studies proved that the electrical activity of the heart can be a candidate of Bio-features for user authentication on mobile IoT devices.

\item  Voice recognition: The voice signal can be used in voice authentication with a characteristic of single-vowel. Kim and Hong \cite{15} used mel-frequency cepstral coefficients and pitch as voice features, and the Gaussian mixture model in the voice authentication process for speaker recognition, as shown in Fig. \ref{fig:Fig2}. Note that voice-based authentication and authorization schemes for mobile IoT devices are vulnerable against attacks that use a pre-recorded voice.

\item  Signature recognition: According to Shahzad et al. \cite{90}, a signature is defined as the conventional handwritten depiction of one's name performed either using a finger. Therefore, existing signature-based authentication and authorization schemes for mobile IoT devices can be divided into three categories, namely, offline, online, and behavior. With the category of offline, authentication and authorization schemes use the form on an image as input signatures. With the category of online, authentication and authorization schemes use the form of time-stamped data points as input signatures. With the category of behavior, authentication and authorization schemes use the behavior of doing signatures with a finger.

\item  Gait recognition: The gait templates can be used for user verification. Based on the biometric cryptosystem (BCS) approach with a fuzzy commitment scheme, Hoang et al. \cite{51} introduced authentication and authorization scheme using gait recognition for mobile IoT devices.

\item  Behavior profiling: Behavior profiling aims at building invariant features of the human behavior during different activities. Frank et al. \cite{39} proposed authentication and authorization scheme using a touchscreen input as a behavioral biometric for mobile IoT devices.

\begin{figure}
\centering
\includegraphics[width=1\linewidth]{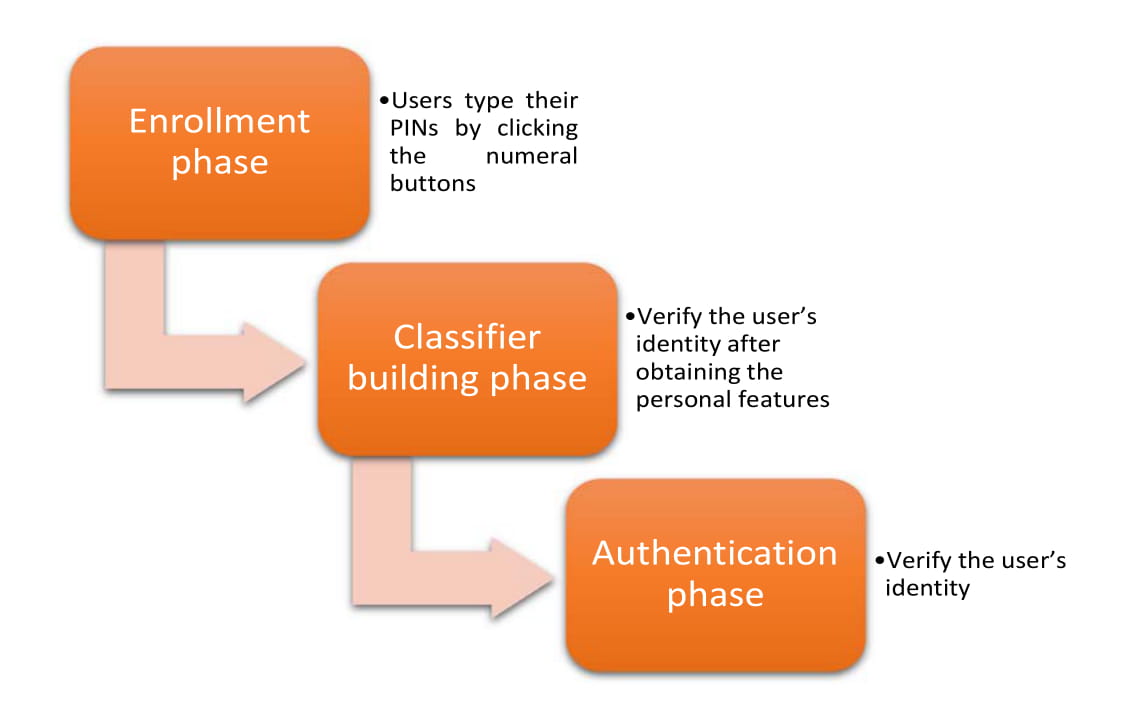}
\caption{A keystroke dynamics-based authentication scheme}
\label{fig:Fig3}
\end{figure}

\item Keystroke dynamics: Existing keystroke-based authentication and authorization schemes for mobile IoT devices can be classified into two types, including, 1) Static, which the keystroke analysis performed only at specific times; and 2) Continuous, which the keystroke analysis performed during a whole session. In order to improve the effectiveness of PIN-based authentication and authorization schemes, Tasia et al. \cite{47} proposed three steps in the keystroke dynamics-based authentication systems, namely, 1) Enrollment step, 2) Classifier building step, and 3) User authentication step, as shown in Fig. \ref{fig:Fig3}.
\begin{figure}
\centering
\includegraphics[width=0.55\linewidth]{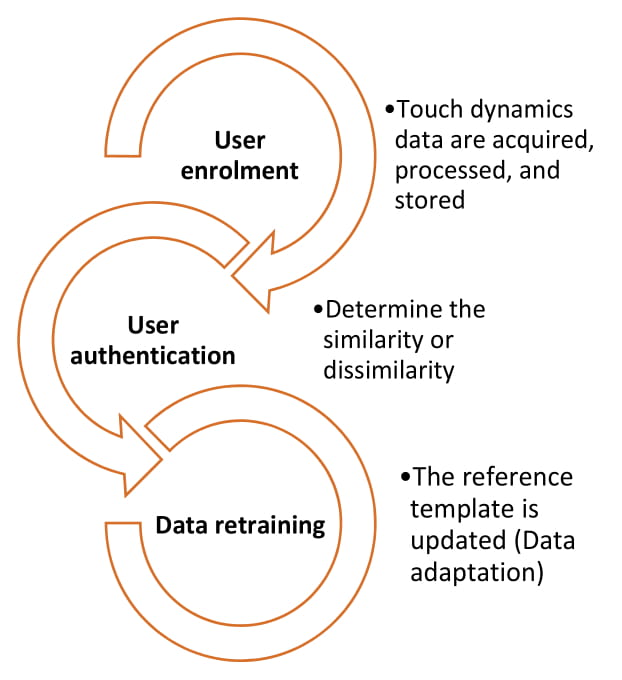}
\caption{An authentication and authorization scheme using touch dynamics for mobile IoT devices}
\label{fig:Fig4}
\end{figure}
\item Touch dynamics: The process of measuring and assessing human touch rhythm on mobile IoT devices is called touch dynamics. According to Teh et al. \cite{6}, the design of a touch dynamics authentication system is performed in three steps, namely, 1) User enrolment step, 2) User authentication step, and 3) Data retraining step, as shown in Fig. \ref{fig:Fig4}.
\begin{figure} [b]
\centering
\includegraphics[width=1\linewidth]{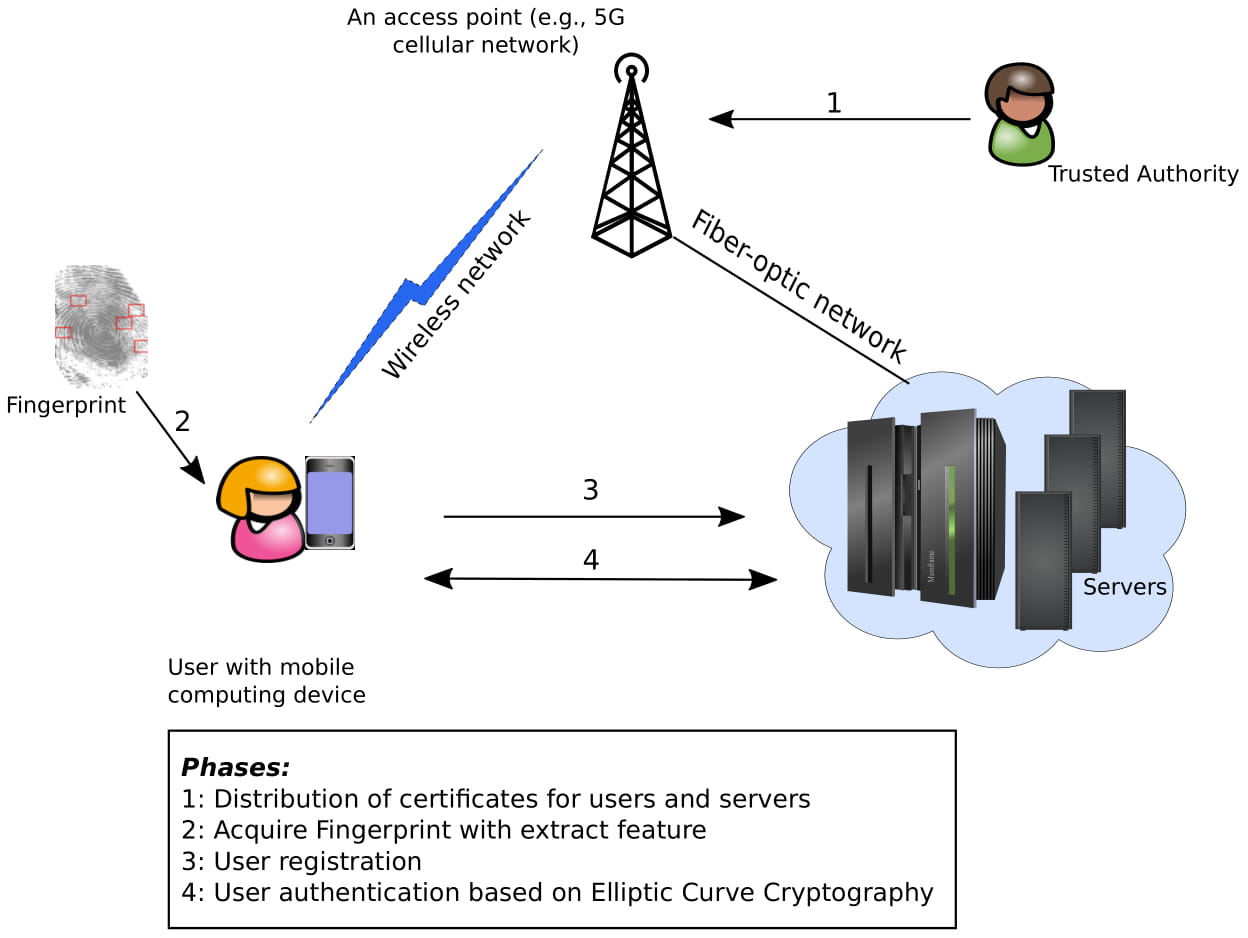}
\caption{Authentication and authorization scheme using fingerprint for mobile IoT devices}
\label{fig:Fig5}
\end{figure}
\item Fingerprint: The fingerprint is used as a bio-key, dynamically to secure a communication channel between client and server after successful authentication on mobile IoT devices. \cite{14,27,32,44}. Currently, authentication and authorization schemes use public key infrastructure framework, such as elliptic curve cryptography, in order to protect the fingerprint biometric, as shown in Fig. \ref{fig:Fig5}.

\item Smart card: According to Li and Hwang \cite{22}, the authentication and authorization for mobile IoT devices using smart cards are one of the simplest and the most effective schemes for IoT authentication compared to traditional password-based authentication schemes. Specifically, the user inputs his/her personal bio-features on mobile IoT device during the registration step. Then, the registration center stores the personal bio-features on the user's smart card.
\item Multi-touch refers to the ability to sense the input simultaneously from more points of contact with a touchscreen \cite{45}. According to Sae-Bae et al. \cite{35}, authentication and authorization for mobile IoT devices using multi-touch gesture are based on classifying movement characteristics of the center of the fingertips and the palm.
\item Graphical password: To withstand dictionary attacks, researchers proposed graphical-based password authentication schemes, which can be classified into two types 1) authentication and authorization using recognition and 2) authentication and authorization using recall.

\begin{figure}
\centering
\includegraphics[width=1\linewidth]{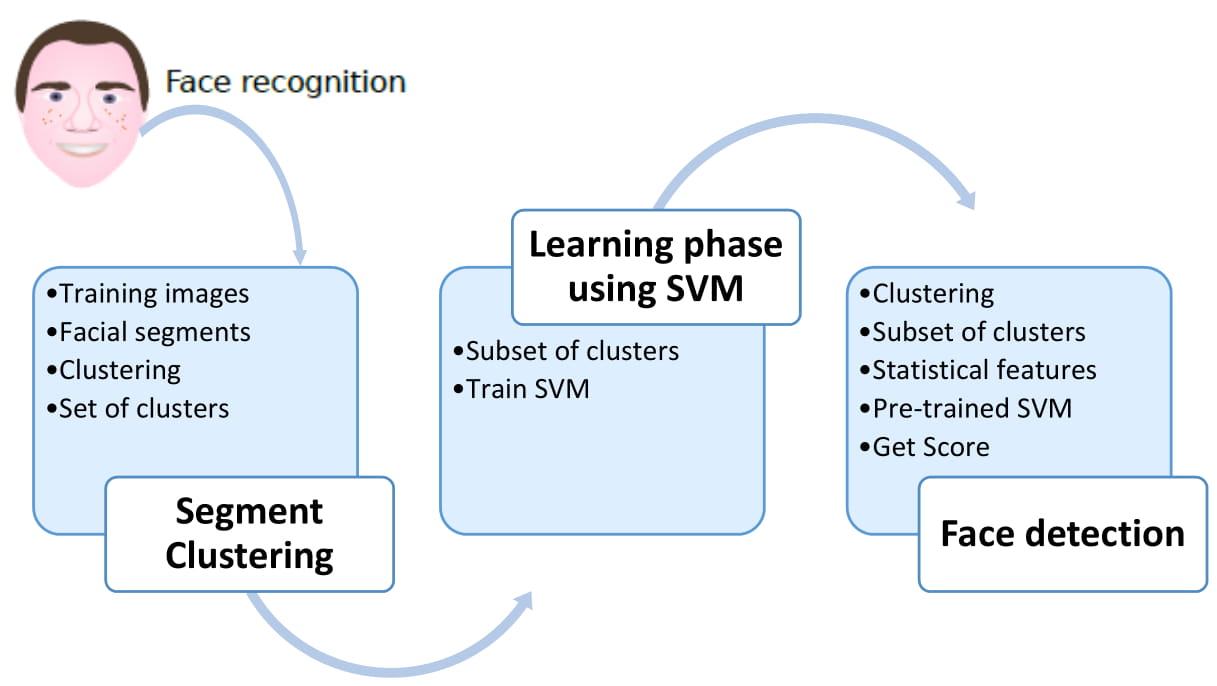}
\caption{A face-based authentication scheme using the Support Vector Machine (SVM)}
\label{fig:Fig6}
\end{figure}

\begin{table*}[!t]
\scriptsize
	\centering
	\caption{Biometric-based authentication schemes for Mobile IoT devices}
	\label{Table:Tab6}
	\setlength{\tabcolsep}{2pt}
		\vspace*{-\baselineskip}
		\renewcommand{\arraystretch}{1.5}
\begin{tabular}{p{0.25in}|p{0.6in}|p{0.6in}|p{1.1in}|p{0.9in}|p{2in}|p{0.7in}} \hline \hline
\rowcolor{black!15}\textbf{Time} & \textbf{Scheme} & \textbf{Method} & \textbf{Goal} & \textbf{Mobile device} & \textbf{Performance (+) and limitation (-)} & \textbf{Complexity} \\ \hline \hline
2007 & Clarke and Furnell \cite{11} & - Keystroke analysis & - Introducing the concept of advanced user authentication & - Sony Ericsson T68;\newline - HP IPAQ H5550; & + Keystroke latency\newline - Process of continuous and non-intrusive authentication & Low \\ \hline 
\rowcolor{black!5}2007 & Clarke and Furnell \cite{13} & - Keystroke analysis & - Enable continuous and transparent identity verification & - Nokia 5110 & + GRNN has the largest spread of performances\newline - The threat model is not defined & High \\ \hline 
2008 & Khan et al. \cite{14} & - Fingerprint & - Introducing the chaotic hash-based fingerprint & - N/A & + Can prevent from server spoofing attack\newline - The proposed scheme is not tested on mobile devices & Low \\ \hline 
\rowcolor{black!5}2010 & Li and Hwang \cite{22} & - Smart card & - Providing the non-repudiation & - N/A & + Can prevent from parallel session attacks\newline - Storage costs are not considered & $10T_H$\newline  \\ \hline 
2011 & Xi et al. \cite{27} & - Fingerprint & - Providing the authentication using bio-cryptographic & - Mobile device with Java Platform & + Secure the genuine biometric feature\newline - Server-side attack is not considered & at FAR=0.1\% , GAR=78.69\% \\ \hline 
\rowcolor{black!5}2012 & Chen et al. \cite{32} & - Fingerprint & - Using only hashing functions & - N/A & + Solve asynchronous problem\newline - Privacy-preserving is not considered & $7T_H$ \\ \hline 
2013 & Frank et al. \cite{39} & - Touchscreen & - Providing a behavioral biometric for continuous authentication & - Google Nexus One & + Sufficient to authenticate a user\newline - Not applicable for long-term authentication & 11 to 12 strokes, EER=2\%--3\% \\ \hline 
\rowcolor{black!5}2014 & Khan et al. \cite{44} & - Fingerprint & - Improve the Chen et al.'s scheme & - N/A & + Quick wrong password detection\newline - Location privacy is not considered\textbf{} & ${18T}_H$\newline  \\ \hline 
2015 & Hoang et al. \cite{51} & - Gait recognition & - Employing a fuzzy commitment scheme & - Google Nexus One & + Efficient against brute force attacks\newline - Privacy model is not defined & Low \\ \hline 
\rowcolor{black!5}2016 & Arteaga-Falconi et al. \cite{61} & - Electrocardiogram & - Introducing the concept of electrocardiogram-based authentication & - AliveCor & + Concealing the biometric features during authentication\newline - Privacy model is not considered. & TAR=81.82\% and FAR=1.41\% \\ \hline 
2017 & Abate et al. \cite{79} & - Ear Shape & - Implicitly authenticate the person authentication & - Samsung Galaxy S4 smartphone & + Implicit authentication\newline - Process of continuous and non-intrusive authentication & EER=1\%--1.13\% \\ \hline 
\rowcolor{black!5}2017 & Khamis et al. \cite{F1} & - Gaze and Touch & - Protect multimodal and authorization on mobile IoT devices & - N/A & + Secure against the side attack model and the iterative attack model\newline - Vulnerable to video attacks & $SSR=68\%\ to\ 10.4\%$ \\ \hline 
2017 & Feng et al. \cite{F2} & - Fingerprints or iris scans & - Introduced a biometrics-based authentication with key distribution & - Google Nexus One & + Anonymity and unlinkability \newline - Interest privacy in not considered & $C_1=8{TE}_{mul}+24T_H$\newline \\ \hline 
\rowcolor{black!5}2017 & Ghosh et al. \cite{F3} & - Fingerprint & - Proposing a near-field communication with biometric authentication & - N/A & + Authentication and authorization for P2P payment\newline - Threat model is not defined & High \\ \hline 
2017 & Mishra et al. \cite{F6} & - Biometric identifier & - Removing the drawback of the Li et al. scheme \cite{F10} & - N/A & + Efficient password change\newline + Off-line password guessing\newline - Location privacy in not considered  & $C_1=8T_H+{TE}_{enc/dec}+2{TE}_{mul}$ \\ \hline 
\rowcolor{black!5}2018 & Li et al. \cite{F4} & - Fingerprint & - Introduced three-factor authentication using fingerprint identification & - N/A & + Quickly detection for wrong password\newline + Traceability of mobile user\newline - Backward privacy is not considered & $C_1=9{TE}_{mul}+2T_e{\rm \ \ }+20T_H$ \\ \hline 
2018 & Yeh et al. \cite{F5} & - Plantar biometrics & - Introduced critical characteristics of new biometrics & - Raspberry PI platform & + High verification accuracy\newline - Threat model is not defined & $RV=83,88\%\ to\ 99,60\%$ \\ \hline 
\rowcolor{black!5}2018 & Bazrafkan and Corcoran \cite{F9} & - Iris & - Use deep learning for enhancing Iris authentication & - N/A & + The iris segmentation task on mobile IoT devices\newline - Privacy preserving is not considered  & $SA=99.3\%$    \\ \hline \hline
\end{tabular}\\
Notations: TAR: True acceptance rate; FAR: False acceptance rate; FPR: False-positive rate; EER: Equal error rate; GAR: Genuine acceptance rate; $T_H$: Time of executing a one-way hash function $SSR$: Shoulder surfing attack rate; $C_1$: Computational cost of client and server (total); ${TE}_{mul}$: Time of executing an elliptic curve point multiplication
; ${TE}_{enc/dec}$: Time complexity of symmetric key encryption/decryption; $T_e$: Time of executing a bilinear pairing operation; $RV$: Accuracy ratio of entity verification; $SA$: Segmentation accuracy.
\end{table*}

\begin{table*}[!t]
\scriptsize
	\centering
	\caption{Threat models and countermeasures}
	\label{Table:Tab1a}
{
\begin{tabular}{p{0.5in}|p{0.7in}|p{1.2in}|p{1.4in}|p{1.4in}} \hline \hline 
\rowcolor{black!15}\textbf{Scheme } & \textbf{Bio-feature } & \textbf{Threat model} & \textbf{Data attacked } & \textbf{Countermeasure } \\ \hline
Khamis et al. \cite{62} & Gaze gestures & - Iterative attacks\newline - Side attacks & - Observe the user several times from different viewpoints & - Multimodal authentication based on combining gaze and touch \\ \hline 
\rowcolor{black!5}Khamis et al. \cite{84} & Gaze gestures & - Shoulder surfing\newline - Thermal attacks\newline - Smudge attacks & - Uncover a user's password & - Multimodal authentication based on combining gaze and touch \\ \hline 
Arteaga-Falconi et al. \cite{61} & Electrocardiogram & - Adversarial machine learning & - Attacking ECG data sensors & - ECG authentication algorithm \\ \hline 
\rowcolor{black!5}Kang et al. \cite{74} & Electrocardiogram & - Adversarial machine learning & - Attacking ECG data sensors & - Cross-correlation of the templates extracted \\ \hline 
Chen et al. \cite{50} & Voice recognition & - Random-guessing attack & - Malicious bystanders try to observe the password of the legitimate user & - Rhythm-based two-factor authentication \\ \hline 
\rowcolor{black!5}Shahzad et al. \cite{90} & Signature recognition & - Shoulder surfing attack\newline - Smudge attack & - Malicious bystanders try to observe the password of the legitimate user & - behavior-based user authentication using gestures and signatures \\ \hline 
Sitova et al. \cite{72} & Behavior profiling & - Population attacks & - Guess the user's feature vector & - Using the notion of guessing distance \\ \hline 
\rowcolor{black!5}Shahzad  et al. \cite{90} & Behavior profiling & - Shoulder surfing attack\newline - Smudge attack & - Spying on the owner when he performs an action & - Authentication scheme based on the gesture and signature behavior\\ \hline 
Khamis et al. \cite{F1} & Touch dynamics & - Side attack model\newline - Iterative attack model & - Spying on the owner when he performs an action & - Multimodal authentication \\ \hline 
\rowcolor{black!5}Ferdowsi and Saad \cite{N1} & N/A & - Eavesdropping attacks & - Extract the watermarked information & - Deep learning algorithm with long short-term memory \\ \hline
Khan et al. \cite{14} & Fingerprint & - Replay attacks, forgery attack and impersonation attack, server spoofing attack & - Replaying of an old login message & - Chaotic hash-based authentication \\ \hline \hline
\end{tabular}
}
\end{table*}

\item Face recognition: Mahbub et al. \cite{70} introduced an authentication and authorization scheme using face recognition, which can be applied for mobile IoT devices. Based on the Support Vector Machine (SVM), the Mahbub et al.'s scheme is based on three steps, namely, 1) Step of segment clustering, 2) Step of learning SVM, and 3) Step of face detection, As shown in Fig. \ref{fig:Fig6}.

\begin{figure} [b]
\centering
\includegraphics[width=0.8\linewidth]{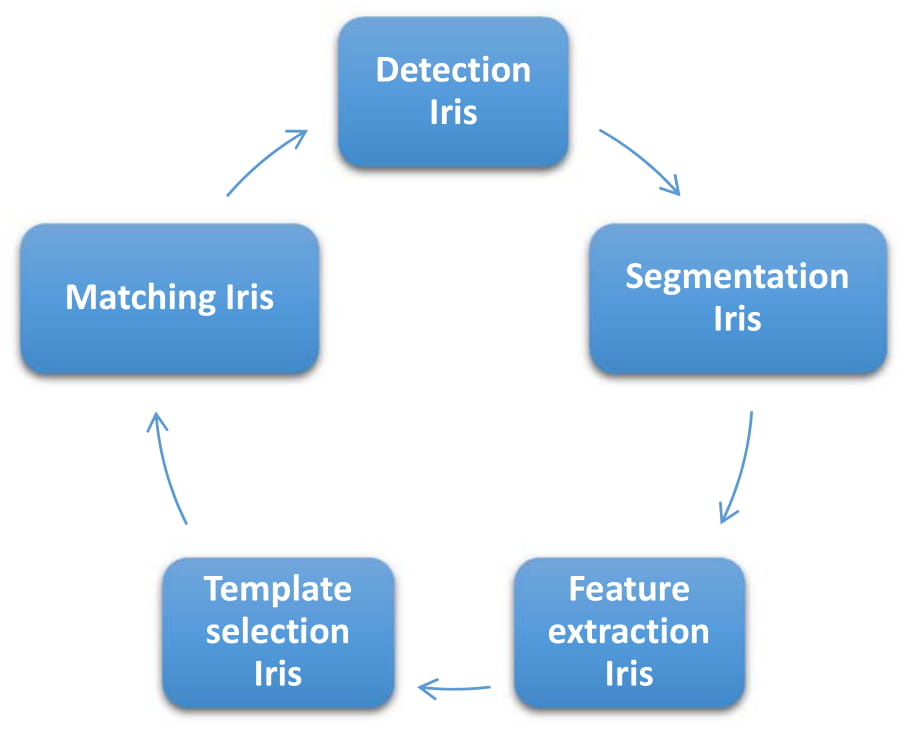}
\caption{A Iris-based authentication scheme}
\label{fig:Fig7}
\end{figure}

\item Iris recognition: Iris-based authentication scheme refers to a comparison with the iris template of the person owning the mobile computing device. This process could be used to unlock a mobile computing device or to validate banking transactions. According to De Marsico et al. \cite{46}, an Iris-based authentication scheme can be repeated in a cyclic process to ensure continuous reidentification, as shwon in Fig. \ref{fig:Fig7}.

\item Rhythmic taps/slides: A rhythm-based authentication scheme refers to user identification by a series of rhythmic taps/slides on a device screen. Chen et al. \cite{50} proposed an authentication and authorization scheme using rhythmic taps/slides, which can be applied for mobile IoT devices. Chen et al.'s scheme is based on two step, namely, 1) Enrollment step and 2) Verification step.

\item Capacitive touchscreen: In order to scan body parts on mobile IoT devices, Holz et al. \cite{52} introduced an authentication and authorization scheme using the capacitive touchscreen. Specifically, the Holz et al.'s scheme appropriates the capacitive touchscreen as an image sensor.

\item Ear Shape: Ear shape-based authentication scheme refers to capturing a sequence of ear images, which are used for extraction of discriminant features, in order to authenticate the users on mobile IoT devices. \cite{79}.

\item  Arm gesture: The arm gesture is usually combined with a physical biometric to authenticate users for mobile IoT devices, e.g. Ear shape \cite{79}.
\end{itemize}

\section{Authentication and authorization schemes for mobile IoT devices using bio-features}\label{sec:bio}

The surveyed papers of Authentication and authorization schemes for mobile IoT devices using bio-features are shown in Table \ref{Table:Tab6}. In addition, threat models and countermeasures are shown in Table \ref{Table:Tab1a}.

The manner and rhythm in which an individual types characters when writing a text message is called keystroke analysis, which can be classified as either static or continuous. For authenticating users based on the keystroke analysis, Clarke and Furnell \cite{11} introduced an authentication and authorization scheme, which is based on three interaction scenarios, namely, 1) Entry of 11-digit telephone numbers, 2) Entry of 4-digit PINs, and 3) Entry of text messages. The Clarke and Furnell's scheme \cite{11} can provide not only transparent authentication of the user, but it is also efficient in terms of FRR and FAR under three types of mobile IoT devices, namely, Sony Ericsson T68, HP IPAQ H5550, and Sony Clie PEG NZ90. To demonstrate the ability of neural network classifiers, the same authors in \cite{13} proposed an authentication framework based on mobile handset keypads in order to support keystroke analysis. The three pattern recognition approaches used in this framework are, 1) Feed forward multi-layered perceptron network, 2) Radial basis function network, and 3) Generalised regression neural network. Therefore, Maiorana et al. \cite{31} proved that it is feasible to employ keystroke dynamics on mobile phones with the statistical classifier for keystroke recognition in order to employ it as a password hardening mechanism. In addition, the combination of pressure and time features is proved by Tasia et al. in \cite{47} that is is among the effective solutions for authentication and authorization.

The passwords have been widely used by the remote authentication schemes, which they can be easily guessed, hacked, and cracked. However, to deal with the drawbacks of only-password-based remote authentication, Khan et al. \cite{14} proposed the concept of chaotic hash-based fingerprint biometrics remote user authentication scheme. Theoretically, the scheme \cite{14} can prevent from fives attacks, namely, parallel session attack, reflection attack, Forgery attack, impersonation attack, DoS attack, and server spoofing attack, but it is not tested on mobile devices and may be vulnerable to biometric template attacks. 

In order to avoid the biometric template attack, Xi et al. \cite{27} proposed an idea based on the transformation of the locally matched fuzzy vault index to the central server for biometric authentication using the public key infrastructure. Compared to \cite{24}, \cite{14}, and \cite{27}, Chen et al. \cite{32} proposed an idea that uses only hashing functions on fingerprint biometric remote authentication scheme to solve the asynchronous problem on mobile devices. In 2014, Khan et al. \cite{44} improved the Chen et al.'s scheme and Truong et al.'s scheme with quick wrong password detection, but location privacy is not considered.

Biometric keys have some advantages, namely, 1) cannot be lost, 2) very difficult to copy, 3) hard to distribute, and 4) cannot be easily guessed. In 2010, Li and Hwang \cite{22} proposed a biometric-based remote user authentication scheme using smart cards, in order to provide  non-repudiation. Without using identity tables and storing password tables in the authentication system, Li and Hwang's scheme \cite{22} can resist masquerading attacks, replay attacks, and parallel session attacks. Authors did not specify the application environment of their scheme, but it can be applied to mobile IoT devices as the network model is not too complicated. Note that Li and Hwang's scheme was cryptanalyzed for several times. 

Touch dynamics for user authentication are initialed on desktop machines and finger identification applications. In 2012, Meng et al. \cite{33} focused on authentication and authorization using user behavioral bio-features such as touch duration and touch direction. Specifically, they proposed an authentication scheme that uses touch dynamics on touchscreen mobile IoT devices. To classify users,  Meng et al.'s scheme performs an experiment with 20 users using Android touchscreen phones and applies known machine learning algorithms (e.g. Decision Tree, Naive Bayes). Through simulations, the results show that  Meng et al.'s scheme succeeds to reduce the average error rate down to 2.92\% (FAR of 2.5\% and FRR of 3.34\%). The question we ask here: is it possible to use the multi-touch as an authentication mechanism? Sae-Bae et al. \cite{35} in 2012, introduced an authentication approach based on multi-touch gestures using an application on the iPad with version 3.2 of iOS. Compared with  Meng et al.'s scheme \cite{33},  Sae-Bae et al.'s approach is efficient with 10\% EER on average for single gestures, and 5\% EER on average for double gestures. Similar to  Sae-Bae et al.'s approach \cite{35}, Feng et al. \cite{36} proposed an authentication and authorization scheme using multi-touch gesture for mobile IoT devices, named FAST, that incurs FAR=4.66\% and FRR= 0.13\% for the continuous post-login user authentication. In addition, the FAST scheme can provide a good post-login access security, but the threat model is very limited and privacy-preservation is not considered.

Arteaga-Falconi et al. \cite{61} introduced the concept of authentication and authorization using electrocardiogram for mobile IoT devices. Specifically, the authors considered five factors, namely, the number of electrodes, quality of mobile ECG sensors, time required to gain access to the phone, FAR, and TAR. Before applying the ECG authentication algorithm, the preprocessing stages for the ECG signal pass by the fiducial point detection. The ECG authentication algorithms is based on two aspects: 1) employing feature-specific percentage of tolerance and 2) employing of a hierarchical validation framework. The results reveal that the algorithm \cite{61} has 1.41\% FAR and 81.82\% TAR with 4$s$ of signal acquisition. Note that ECG signals from mobile IoT devices may be affected by noise due to the type of motion and signal acquisition, as discussed by Kang et al. \cite{74}. However, the advantage of using ECG authentication is concealing the biometric features during authentication, but it is a serious problem if privacy preservation is not considered. 

\section{Future Directions}\label{sec:future}
Several challenges still remain that opens interesting research opportunities for future work, including, doppler radar, vocal resonance, mobile malware threats, and adversarial machine learning.

\subsection{Doppler radar}
A team of researchers at Buffalo University, led by Wenyao Xu, developed a system that exploits a Doppler radar capable of "reading" the human heart! It works roughly like any other radar, emitting microwaves and analyzing the return signal in order to detect changes in motion \cite{lin2017cardiac}. As scientists say, the process of identifying a person through the method takes about eight seconds, and radar power is just 5 milliwatts - which means that radiation is not dangerous to the body. This method can be a basis for future biometric systems that can be fast, efficient and recognize unique characteristics of the human body.

\subsection{Vocal Resonance}
In \cite{liu2018vocal}, the authors proposed using vocal resonance, that is, the sound of the person's voice as it travels through the person's body. Vocal resonance can be used as a passive biometric, and it achieves high accuracy in terms of identification and verification problems.  It is a method  that is suitable for devices worn on the chest, neck, or initially but could also be used in the near future for recognizing any device that a user posses.

\subsection{Mobile malware threats against biometric reference template}
In 2016 \cite{ref1,ref2}, an Android malware succeeded in bypassing the two-factor authentication scheme of many banking mobile applications that are installed on the user's mobile device. The malware can intercept two-factor authentication code (i.e., verification code sent through SMS), and forward it the attacker.  In case of biometric-based authentication, this threat can be evolved to access the biometric reference template, which are stored at the mobile device, and send it to the attacker. One research direction to prevent this kind of attacks is to employ policy-enforcement access control mechanisms that are appropriate for resource-constrained mobile devices.

\subsection{Adversarial machine learning against biometric-based authentication schemes}
Some biometric-based authentication mechanisms, and especially behavioral-based ones, use machine learning techniques for extracting features and building a classifier to verify the user's identity. Adversarial machine learning aims to manipulate the input data to exploit specific vulnerabilities of the learning algorithms. An adversary using adversarial machine learning methods tries to compromise biometric-based authentication schemes and gain illegal access to the system or the mobile device. The future research efforts should focus on dealing with this kind of threats.

\subsection{Machine learning and blockchain-based authentication}
The blockchain technology is being used in different application domains beyond the cryptocurrencies, e.g., SDN, Internet of Things, Fog computing, etc.\cite{ferrag2018blockchain}. To developing a machine learning and blockchain-based solution for authenticating mobile IoT devices, we have to take in mind the specific requirements of the blockchain, e.g., 1) when IoT data needed to be checked by the IoT entities without any central authority, 2) the ledger copies are required to be synchronized across all of the IoT entities $\cdot$ etc. In addition, the vulnerabilities of the peer-to-peer blockchain networks during the authentication need to be considered, including, private key leakage, double spending, transaction privacy leakage, 51\% vulnerability, and selfish and reputation-based behaviors. Hence, the machine learning-based authentication schemes using the blockchain technology should be investigated in the future.

\subsection{Developing a novel authentication scheme}
For developing a novel authentication scheme for mobile IoT devices using bio-features, we propose the following six-step process:

\begin{enumerate}
\item Definition of IoT network components (Cloud computing, Fog computing, IoT devices, ...etc),
\item Choose the threat models (e.g., iterative attacks, shoulder surfing attacks, thermal attacks, smudge attacks, eavesdropping attacks),
\item Choose the bio-features (e.g., face, eyes, fingerprints-palm, electrocardiogram, signature, voice, gait, keystroke, ...etc).
\item Choose the machine learning and data mining methods (unsupervised, semi-supervised, or supervised),
\item Proposition of the main steps (e.g., Enrollment steps, Classifier building step, and user authentication step),
\item Evaluate the scheme's performance using classification metrics, including, TAR, FAR, FPR, EER,...etc.
\end{enumerate}
\section{Discussion}\label{sec:discus}
There is a big discussion regarding the use of biometric characteristics of the users from new systems or technologies.  Biometric technology can be used to protect privacy, since only a minimum amount of information is required to determine whether someone is authorized, for example, to enter a specific area. On the other hand, since biometrics can reveal sensitive information about a person, controlling the usafe of information may be tricky, especially now that the technology has reached the stage of being applied in mobile devices which can be easily lost or stolen \cite{royakkers2018societal}. Those who are against the use of such features raise concerns about how these data are going to be used. These   concerns could be mitigated by making clear to people that their data is only stored for a limited time, and explaining who will process this data and for what purposes \cite{oostveen2014non}. 
To that sense, the General Data Protection Regulation (GDPR) for European Member States addresses biometric data storage and processes in terms of data protection and privacy. EU countries are affected including the UK and all companies that store or process data of EU citizens. 
On the other hand, in the United States, there is no single comprehensive federal law regulating the collection and processing of biometric data. Only three states Washington, Texas, and Illinois, which have a biometric privacy law in spite that US regulators are also increasingly focusing on the protection of biometric data. Moreover,  In August 2017, India's supreme court decision about a landmark case that named privacy a "fundamental right"showcased that biometric data protection is top on regulators' agenda.

Except from data use issues, general terms such as $computer fear$ and $technophobia$ also provide established accounts of individuals resistance to use new and unfamiliar information technologies, especially for elder people \cite{selwyn2003apart}. Moving one step further, companies that produce applications or methods that use biometric characteristics must comply with a code of ethics or a consistent legal framework governing this kind of data collection which is still absent. For that reason IEEE P7000, is the first standard IEEE is ever going to publish on ethical issues in system design in the next couple of years \cite{spiekermann2017ieee}.

\section{Conclusion}\label{sec:con}
In this article, we have presented a comprehensive literature review, focusing on authentication and authorization for mobile IoT devices using bio-features, which were published between 2007 and 2018. We presented the machine learning and data mining algorithms used by authentication and authorization schemes for mobile IoT devices, including, unsupervised, semi-supervised, and supervised approaches. We reviewed all the Bio-features used by authentication and authorization schemes for mobile IoT devices. We presented the pitfalls and limitations of the existing authentication and authorization schemes for mobile IoT devices.  Several challenging research areas (e.g., doppler radar, vocal resonance, mobile malware threats, adversarial machine learning, machine learning and blockchain-based authentication) will open doors for possible future research directions for mobile IoT devices.

\section*{Conflicts of Interest}
The authors declare that they have no conflicts of interest.

\bibliographystyle{IEEEtran}
\bibliography{ferragmobile}

\end{document}